\documentclass[twocolumn,showpacs,preprintnumbers,amsmath,amssymb]{revtex4}

\usepackage{graphicx}
\usepackage{dcolumn}
\usepackage{bm}

\begin{document}
	
\title{Implications of the geometric representation of the early universe wave function }
	
\author{Bohdan Lev$^{a,b,c}$}

\affiliation{ a). Bogolyubov Institute for Theoretical Physics, NAS of Ukraine, Metrolohichna 14-b, Kyiv 03143, Ukraine, b). Condensed Matter Physics Department, J. Stefan Institute, Jamova 39, 1000 Ljubljana, Slovenia. c). Faculty of Mathematics and Physics, University of Ljubljana, Jadranska 19, SI-1000, Ljubljana, Slovenia,e-mail bohdan.lev@gmail.com}
	
	\date{\today}
	
\begin{abstract}
	
The main goal of this article is to present an algebraic approach to describing the birth and start evolution of the universe. In such an approach, it is possible to use the nature of supersymmetry in terms of the geometric representation of the wave function and propose a mechanism of spontaneous symmetry breaking of the excitations of the universe with different degrees of freedom. On this basis, it is possible to explain the origin of dark energy and matter and explain the baryonic asymmetry of the universe.

\end{abstract}
\maketitle
	
\textbf{Introduction} 
The problem of describing the early universe is far from being solved. Modern ideas (rather hypotheses) about the reason for the formation of the current state of the universe assume the instability of the vacuum in the presence of a fundamental scalar field associated with the quantum nature of matter \cite{LIN}. The reasons and physical mechanism for the origin of this field, and therefore the origin of the universe, have remained open to debate for many years. Furthermore, the physical nature of dark matter remains uncertain, and the existence of dark energy, although it has some physical explanations, is in no way related to dark matter. It seems more natural to look for the causes of these physical formations on the basis of a unified approach to the description of the early universe.

The general issue of cosmology is the determination of the physical and, with it, the geometric nature of the fundamental field. The main assumption of this article will be that the fundamental field can have not only a scalar nature, but also other geometric representations. It is natural that its geometric characteristics should derive from the space that was formed as a result of the distribution of the born substance. In terms of relevant physical characteristics, the Clifford number is the most suitable for this definition \cite{Lev10}-\cite{Lev6}. Special aspects of this geometric representation will be used in this article. The current approach to quantum gravity \cite{Opp},\cite{Wolk} suggests that probabilistic quantum theory must be transformed into a geometric form in order to combine it with general relativity, which represents it.

As it was shown earlier \cite {con} - \cite {cin}, the application of the Clifford algebra covers all standard functions of quantum mechanics and provides \cite {lev} a unifying basis for the physical knowledge including the general relativity and electromagnetism. When we use the Clifford algebra in the scheme of quantum mechanics \cite {con},\cite{con1} we should not ignore the specifics of this formulation. Actually, in this case we obtain a quantum-mechanical theory that provides only an algebraic structure and does not contain any further specific requirements. It is possible to show \cite {con} - \cite {cin} that Clifford's algebraic formalism is completely equivalent to the traditional approach to quantum mechanics.

An \cite{Lev5},\cite{Lev7} approach was previously proposed to describe the origin and evolution of the universe in terms of first principles statistical mechanics and quantum field theory. With this approach, it is possible to answer the question about the probability of born of such a field, as well as about its physical and geometric nature. In terms of some physical entity, which has a simple geometric interpretation using the well-known mathematical apparatus of Clifford's algebra, it is possible to describe the behavior of the early universe and explain the reasons for the appearance of dark matter and energy. This task will be the main goal of the presented article.

The main purpose of this paper is to present a new results of a natural approach to the geometrical description of the birth and evolution of the universe. The wave function of universe as a natural fundamental field will be represented by a Clifford number with transfer rules that have the structure of the Dirac equation for any manifold. In terms of such a geometric representation of the wave function, it is possible to explain the nature of supersymmetry and propose a mechanism for the spontaneous symmetry breaking of the excitations of the universe with different degrees of freedom, and to explain the necessity of the existence of dark matter and energy. 

According to the principle of energy minimization and entropy increasing , was derive the necessary Hamiltonian that can describe the early universe in terms of a geometric interpretation of the wave function.  On this basis, it is possible to explain the asymmetry between the Bose and Fermi degrees of freedom of the universe and obtain non-standard conditions of spontaneous symmetry breaking with condensation of fields of different tensor dimensions. This opens up the possibility of another interpretation of quantum phenomena in the early universe. 

\textbf{Wave function of the Universe}

The most acceptable mathematics structure  is the Clifford algebra. This algebra is a vector space over the field of real numbers \cite{fro}. In this algebraic structure there are ideals that can be obtained by multiplying an isolated element from the right or left by the elements of the ring \cite{hil}, \cite{kas}. The ideal after this procedure is simple Dirac spinors in the standard approach. Thus, the Clifford algebra representation contains more information about the physical properties than the spinor representation \cite{Dor} and can be extended  to  description of the origin and evolution of the universe.

As show the previous study \cite{con}-\cite{Hor} - the quantum mechanics is emerging from the mathematical structure with no need to appeal to an external Hilbert space of wave functions . The Dirac equation as a transfer rule for the wave function on any manifold has a hidden geometric structure and can be used as interpretation for the quantum mechanics \cite{lev},\cite{Lev10}, \cite{con1}. In this sense,is important the geometric representation for the generator of electromagnetic transformations, as well as for the electroweak calibration group of the Weinberg-Salam model \cite{Jos}. This geometric structure also helps reveal more closer connections with the classical theory, than was believed until now. Clifford's space-time algebra illustrates another form of the same wave equation. Tensors built from Dirac spinors look different and it is easier to obtain the self relation between these tensors, and in this sense they differ from all classical physics \cite{Dav}. Based on geometric representations, it is easier to get another interpretation of the obtained results.

The starting point is that our space-time is four-dimensional. We use the basic idea of correspondence between spinor matrices and elements of the external algebra and thus define the state space in terms of space of representations of the Clifford space-time algebra $Cl_{1,3}$. It may be assumed that each elementary formation at an arbitrary point may be described in terms of a Clifford number. Then the wave function of an arbitrary excitation may be represented by a complete geometric object the sum of probable direct forms of the induced space of the Clifford algebra \cite{Lev10}, \cite{Lev7}, \cite{Lev6}. In this case the full geometric entity  may be written in terms of the direct sum of a scalar, a vector, a bi-vector, a three-vector, and a pseudo-scalar, i.e, $\Psi=S\oplus V\oplus B\oplus T\oplus P$, that is given by
\begin{equation}
\Psi=\Psi_{0}\oplus\Psi_{\mu}\gamma_{\mu}\oplus\Psi_{\mu \nu
}\gamma_{\mu}\gamma_{\nu}\oplus\Psi_{\mu\nu\lambda}\gamma_{\mu}\gamma_{\nu}\gamma_{\lambda}\oplus\Psi_{\mu \nu \lambda \rho}\gamma_{\mu}\gamma_{\nu}\gamma_{\lambda}\gamma_{\rho} \label{71}
\end{equation}
In the reverse order of the composition, we can change the direction of each basis vector and thus obtain $\bar{\Psi} =S\ominus V\oplus B\ominus T\oplus P$.
Another element of symmetry is the change of multiplication of base vectors to inverses in the representation of Clifford numbers, which turns it into $\tilde{\Psi}=S \oplus V\ominus B\ominus T \oplus P $. Along with the symmetry elements, the ring structure is satisfied using the direct tensor product in the symbolic notation given as
\begin{equation}
\Psi\Phi=\Psi\cdot\Phi+\Psi\wedge\Phi, \label{61}
\end{equation}
where $\Psi\cdot\Phi$ is an inner product or convolution that decreases the number of basis vectors and $\Psi\wedge\Phi$ is an external product that increases the number of basis vectors. If we multiply each Clifford number by a fixed matrix $u$ with one column, where the first element is one and all others are zero, we get a Dirac bispinor with four elements. Using this column, one can reproduce the spinor representation of each Clifford number. A complex conjugate bispinor can be obtained by multiplying the same Clifford number by a string $u^{+}$ whose first element is one and all others are zero $u u^{+}=1 $. Between such bispinors and elements of the external algebra there is a complete correspondence - isomorphism.

Now let us determine the rule of comparing two Clifford numbers in different points of the manifold. For this purpose we have to consider the deformation of the coordinate system and the rule of translations on different manifolds. An arbitrary deformation of the coordinate system may be set in terms of the basis deformations $e_{\mu}=\gamma_{\mu}X$, where $X$ is the Clifford number that describes arbitrary changes of the basis (including arbitrary displacements and rotations), that do not violate its normalization, i.e., provided $\tilde{X}X=1$. It is not difficult to verify that $e^{2}_{\mu}=\gamma_{\mu}\tilde{X}\gamma_{\mu}X= \gamma^{2}_{\mu}\tilde{X}X= \gamma^{2}_{\mu}$, and this relation does not violate the normalization condition \cite{kas}. An arbitrary particle must be represented by a mathematical object with
corresponding transformation properties during rotations and translations. The particles presented here have properties of the spinor transformation \cite{Had}, so any field theory that attempts to model for spin behavior will necessarily use spinor fields as well as their richer representation.

Now, for an arbitrary basis, we may set, at each point of the space, a unique complete linearly independent form as a geometric entity that characterizes this point of the manifold. If this point of the manifold is occupied,  then its geometric characteristics may be described by the coefficients of this representation. A product of arbitrary forms of this type is given by a similar form with new coefficients, thus providing the ring structure. This approach makes it possible to consider the mutual relationship of fields of different physical nature \cite{lev},\cite{kle}. To determine the characteristics of the manifold as a point function implies to associate each point of the set with a Clifford number and to find its value. If this function is differentiable with respect to its argument, we may introduce the differentiation operation. To define a transfer operation on an arbitrary manifold, we have to define a derivative operator, e.g., as given by $ D = \gamma_{\mu} \frac {\partial}{\partial x_{\mu}} $ where $ \frac {\partial} {\partial x_{\mu}} $ represents the change along the curves passing through a given point in the space. The action of this the operator for any Clifford number may be presented as
\begin{equation}
D\Psi=D\cdot\Psi+D\wedge\Psi
\end{equation}
where $ D \cdot \Psi $ and $ D \wedge \Psi $ may be referred to as the "divergence" and the "rotor" of the relevant Clifford number. Within the context of the definition of a differentiated variety, it is not enough to have one non-special coordinate system covering a variety whose topology differs from the topology of an open set in the Euclidean space

Furthermore, by ascribing a given geometric interpretation to the wave function, we may obtain correct transfer rules for an arbitrary variety \cite{lev} and obtain new results concerning the geometric nature of the wave function. For the wave function as a geometric entity, we  may write the first structure equation in the standard form
\begin{equation}
\Omega=d\Psi-\omega \Psi=m \Psi \label{81}
\end{equation}
where covariant derivation $\Omega$ from wave function is propositional from same wave function. In our approach, the dynamic equation for the wave function is represented as a rule of parallel transfer on an arbitrary manifold where $\omega$  is the connectivity of the manifold and have the same geometric representation of the Clifford number as the wave function \ref{71}. For a complete group of linear transformations $ \Psi'=\Psi X$ , where $X$ defines the mapping elements and satisfies the condition $\tilde{X}X=1$, the calibration transformation for the connectivity $\omega$ is given as $ \omega'=X \Phi \tilde{X}+Xd\tilde{X}$.

 The equation \ref{81} reproduces the form of the Dirac equation but with fuller meaning than in the spinor representation \cite{kas}. Such Dirac equation for the wave function  may be obtained by minimizing the action constructed from the geometric invariant 
\begin{equation}
S=\int d\tau \left\{\Omega\widetilde{\Omega}+m^{2}\Psi\widetilde{\Psi}\right\}\label{action}
\end{equation}
The Lagrange multiplier $m^{2}$ provides the normalization condition for the wave function $\int d\tau \left\{\Psi\widetilde{\Psi}\right\}=1$. The above action is non-degenerate for the solution of the Dirac equation, unlike the standard approach. In our approach, the dynamic equation for the wave function is presented as a parallel transfer rule on an arbitrary manifold.

For further consideration, we need to define the scalar product $\Psi\widetilde{\Psi}$. Let's return to the the general representation of the wave function from \ref{71} equation. In the context of the definition of it \cite{kas},\cite{Dav} can be seen that the wave function in the general case can be written as the sum of even and odd part $\Psi=S\oplus B\oplus P\oplus V\oplus T$, where $S\oplus B\oplus P=q +i\acute{q}=Q$ is the biquaternion ($q$ and $\acute{q}$ are quaternions), and the sum $ V\oplus T=V+i\acute{V}$ contains a vector and a trivector or a pseudovector. In this representation, the scalar product takes the form \cite{Hes}, \cite{kas}:
\begin{equation}
\rho=\Psi\tilde{\Psi}= (q+i\acute{q})(\tilde{q}-i\tilde{\acute{q}})+(V+i\acute{V})(\tilde{V}-i\tilde{\acute{V}})=\rho \exp i \beta \label{51}
\end{equation}
where $\rho$ can have a physical interpretation as the density of the probability of finding a particle at an arbitrary point of the manifold. The physical content of the entered quantities can be understood from the following. The product of the wave function by its conjugate has two parts - the inner and the outer. Let's insert a unit between them represented by projections $u$ and apply that projection to each wave function. After such manipulation, the wave function is transformed into a bispinor and the conjugate wave function into a conjugate bispinor
\begin{equation}
\Psi u=
\left( \begin{array}{cc}
B&  \\
F&  
\end{array}\right),    u^{+}\tilde{\Psi}=(F^{*},B^{*})
\end{equation} 
where $F$ and $B$ spinors. As a result, the inner product of wave functions is transformed into a scalar product $S_{+}=\Psi \cdot \tilde{\Psi}= (B^{*}F +F^{*}B) $ and the outer product into a "vector" product of bispinors $S_{-}=\Psi \wedge \tilde{\Psi}=i( B^{*}F -F^{*}B $ \cite{Dav},\cite{hes}.  In such presentation $\rho =\sqrt{S^{2}_{+}+S^{2}_{-}}$ and $\tan \beta=\frac{S_{-}}{S_{+}}$.  We  will return to the definition of the physical content of the angle $\beta$ later. In this way, it becomes possible to describe the intermediate states of many bodies, since the form of the ensemble of wave functions will be analogous to \cite{tre}- \cite{Sch}. Such a wave function can play the role of a fundamental field for the early universe. At the same time, it is not very important which quantum or classical interpretation we attribute to it.

\textbf{Cosmological model in the geometrical presentation}

Let's start with the physical interpretation of the geometric nature of the fundamental field. We do not know in advance what the universe was born from, but we do know what it is made of. These are particles and fields that have the same geometric interpretation. For this reason, we must introduce a physical quantity that can reproduce these properties after the birth of the universe. The geometry of space can be created only after the distribution of particles and fields. That is, the characteristics of both possible particles and fields, as well as the properties of a possible variety, where its evolution will be considered, should be presented. What we now observe may be created from an entity that has both corpuscular and wave properties.Such a geometric entity corresponds to the Clifford number representation, where particles and fields characteristics are automatically included at the same time. Based on such assumptions, we will try to change the emphasis in the previously obtained results and give an explanation to some new physical facts.

As was proposed in articles \cite{Lev10},\cite{Lev11} the early universe can describe in the terms geometrical presented wave function. This wave function play role fundamental field,  which fully describe further behavior of universe.   In the case of spontaneous generation of an additional field in vacuum, the ground state energy of the "new" vacuum for fields of different nature should be lower than the ground state energy of the "initial" vacuum \cite{Lev6}. We assume that occurrence in vacuum of the fundamental field that is generated spontaneously and interacts with the fluctuations of all other fields may be described in terms of the Clifford number \cite{lev}. The probable stationary states of the fundamental field are generated by the multiplicative noise produced by the nonlinear self interaction with fluctuations of this field. The generator of these quantum fluctuations is the vacuum itself for each point of the Planck size on the manifold.
	
This model differs from the widely studied scenario of stochastic inflation of the universe \cite{LIN}, which takes into account fundamental field fluctuations, but does not take into account fluctuations of the unstable vacuum. Internal fluctuations generate stochastic behavior of the system, which can cause changes in its steady state. The most essential point is that the fundamental field is a Clifford number and not a scalar, and contains all the geometrical characteristics of the possible space that can be formed as a result of the emergence of matter. 
	
Can start with the assumption that the  transition from the "initial" vacuum to the new state of vacuum generates a new non-zero entity. The new entity generates the "new" vacuum different from the "primary" vacuum for any field of an arbitrary geometric characteristic that may appear. The resulting field must reduce the energy of the "new" vacuum with respect to the energy of the "primary" vacuum. Therefore, the energy  of the ground state of the "new" vacuum may be presented through 
\begin{equation}
E=E_{v}-\frac{\mu_{0} ^{2}}{2} \Psi \widetilde{\Psi} \label{11}
\end{equation}
where the coefficient $\mu_{0} ^{2}$ describes the coupling of the new field and the "primary" vacuum, i.e., the self-consistent interaction of the new field with the probable fluctuations that may exist in the "primary" vacuum. Here we have to make two remarks. The first one concerns the decrease in the initial energy of the ground state with the appearance of a new field, and the second one is related to the coupling coefficient that is now positive, so that explanations of the appearance of such a sign used in the standard approach are not needed. The energy of the new state may be presented in the form

If we want to describe the evolution of the system $\langle out| \exp iHt |in \rangle$, we still need to average all probable fluctuations with which the new field can interact. For this purpose it is sufficient to present the coupling coefficient in the form $\mu_{0} ^{2}=\mu ^2+\xi $, where $ \langle \xi(t)\xi(0) \rangle=\sigma ^2$ and $\sigma ^2$ is the dispersion and after that to carry out averaging
\begin{eqnarray}
		\langle out| \exp \frac{i}{h}Ht |in
	\rangle \sim  \\
\langle \int D \Psi \int D\xi \exp \frac{i}{h}\left\{ E_{v}-\frac 12\mu ^2\Psi \widetilde{\Psi}+\frac 12 \xi\Psi \widetilde{\Psi}  +\frac{\xi ^2}{\sigma ^2}\right\}t|in\rangle \\
\sim \sqrt{4\pi \sigma}\int D \Psi \exp \frac{i}{h}\left\{ E_{v}-\frac 12\mu ^2\Psi \widetilde{\Psi}+\frac{\sigma ^2}{4} (\Psi \widetilde{\Psi})^2 \right\}\nonumber
\end{eqnarray}
after integration over fluctuation fields yields. This implies that we have a new system with the effective energy (averaged over the fluctuations of  other fields) given by
\begin{equation}
	E==E_{v}-\frac 12\mu ^2\Psi \widetilde{\Psi}+\frac{\sigma ^2 }{4}(\Psi \widetilde{\Psi})^2 =E_{v}+V(\Psi)\label{14}
\end{equation}
where introduce the effective potential $V(\Psi)=-\frac 12\mu ^2\Psi \widetilde{\Psi}+\frac{\sigma ^2 }{4}(\Psi \widetilde{\Psi})^2 $ of the fundamental field in the geometric interpretation, that reproduces the well-known expression for the energy of the fundamental scalar field but with the nonlinear coefficient determined by the dispersion of fluctuations. This potential reproduces all the consequences of the behavior of the new vacuum in the standard approach and is responsible for the dark energy of the theory of gravity in the geometric interpretation. This implies that with no field $\Psi=0$, $E=E_{v}$ while for $ \Psi \widetilde{\Psi}=\frac{\mu ^2}{\sigma^{2}}$ the expression for the effective ground state energy of the "new" vacuum reduces to $ E=E_{v}-\frac{\mu ^4}{4 \sigma^{2}} $. As follows from the latter relation, the energy of the "new" vacuum is lower than the energy of the primary vacuum, i.e. the phase transition results in the formation of a new vacuum ground state. If $\sigma^{2}$ tends to infinity, then the energy of the new state tends to the initial energy of the ground state. If the energy of the initial state is equal to $E_{v}=\frac{\mu ^4}{4 \sigma^{2}}$  this relation can be applied to estimate the maximum dispersion of field fluctuations, provided that the initial vacuum temperature is given by this relation.
	
\textbf{Supercharge in the geometric representation}
	
Now we propose a slightly different scenario for the birth of the Universe based on the representation of its wave function as a geometric entity, a Clifford number with an appropriate physical interpretation. An additional field is required for the emergence of the matter, whose spontaneous excitation leads to the emergence of elementary particles. Solving the question of the impact of the early supersymmetric quantum cosmological era on current cosmological observations was the purpose of the paper \cite{Mon},\cite{Mar}. Prospects of quantum cosmology are presented in a comprehensive review \cite{Mon1}. In our case, such a field is the wave function $\Psi$ in various tensor representations, that is, it has all possible tensor representations with the dimensions of the created space. That is, the geometry is embedded from the very beginning in the characteristics of the point of the manifold on which we describe it. The manifestation of geometry still requires the birth of particles, the distribution of which creates it.
	
It was previously proved that in the presence of a spontaneously generated fundamental field, the energy of the vacuum state for any other field is lower than the energy of the ground state of the primary vacuum, and that the energy of the fundamental field is affected by its nonlinear interactions with fluctuations of physical fields of different nature. To avoid the problem of the influence of the gravitational field on the evolution of the universe at the stage of spontaneous nucleation of the fundamental field, we note that the energy of the primary vacuum is not contained in the Einstein equation, and the dynamics of the universe is determined only by the potential energy of the fundamental field that produces the matter. The distribution of the matter, in turn, determines the geometry.

According to Dirac's theory we may move from the classical Poisson brackets to the quantum ones and rewrite the Hamiltonian in terms of the secondary quantization, where instead of classical geometric representations of the wave function we introduce the operators of birth and annihilation of quanta of this field. In the operator form the Hamiltonian of the Universe may be written as:
\begin{equation}
	H=E_{v}-\frac 12\mu ^2\hat{\Psi}^{+}\hat{\Psi}+\frac{\sigma ^2 }{4}\hat{\Psi}^{+}\hat{\Psi} \hat{\Psi}^{+}\hat{\Psi} \label{37}
\end{equation}
For the field operators of the general form thus introduced, the commutation relations may be unusual. We note that this field has by definition representation of both boson and fermion fields that should include probable transformations of bosons into fermions and vice versa. For mathematical justification of the proposed ideas, let's return to Clifford's numbers. As \cite{Lun}- \cite{Gu} shows early, the scalar product of Clifford numbers $\hat{\Psi} \cdot \hat{\tilde{\Psi}}=S_{+}= (\hat{F^{+}}\hat{B}+ \hat{F}\hat{B^{+}})$  can be represented as the product of a bispinor on its conjugated bispinor. From previous consideration we can assume that bispinor $F$ describes particles with half spin and bispinor $B$ particles with whole spins. $\hat{B}^{+}$ and $\hat{B}$ denote Bose creation and annihilation operators and let $\hat{F}^{+}$ and $\hat{F}$ denote Fermi creation and annihilation operators with the (anti)commutation relation $[\hat{B},\hat{B}^{+}]=\{\hat{F},\hat{F}^{+}\}=1$, $[\hat{B},\hat{F}^{+}]=[\hat{F},\hat{B}^{+}]=\hat{F}^{2}=(\hat{F}^{+})^{2}=0$. As it was shown earlier, the value of the density $\rho$ plays the role of the supercharge $S_{+},S_{-}$ that describes the intensity of the transfer between different degrees of freedom in the general representation of the wave function. As shown earlier in the geometric representation there is another supercharge $S_{-}= (\hat{F^{+}}\hat{B}-\hat{F}\hat{B^{+}})$ which also needs to be considered. To clarify the main essence of the task of this article, it is sufficient to limit ourselves to only the scalar part of the Hamiltonian (energy). Even in such an abbreviated version of the representation, it is possible to obtain a non-trivial interpretation of the behavior of the early universe.Now it is not difficult to see that the Hamiltonian may be written in the form supersymmetry theory as given by
\begin{equation}
	H=E_{v}-\frac 12 \mu ^2 S_{+}+\frac{1}{4}\sigma ^2 S^{2}_{+} \label{34}
\end{equation}
where the square of the supercharge is $S^{2}_{+}=(\hat{B}^{+}\hat{B}+\hat{F}^{+} \hat{F}^{+})=n_{B}+n_{F}$ and presents the total number of bosons and fermions. New commutation relation can be present in the form:$[\hat{H},S^{2}_{+}]=0$ It is obvious that all the elements of the supersymmetry with the commutation relations  $[\hat{S_{+}},\hat{S^{2}_{+}}^{2}]=0 $ are contained in the presented form, where the individual parts of the Hamiltonian are associated with the integrals of motion and are preserved both separately and together. The interpretation of the supercharge in our case is that the geometric representation of the wave function provides a possibility to consider a physical mixture of bosons and fermions, and the charge itself describes the probable transformation of particles into each other. That is, the initial wave function describes a mixed state of bosons and fermions with probable mutual transformations of individual components. If we calculate the partition function \cite{Kap}
	\begin{equation}
	Z=Tr \exp\beta [\frac 12\mu^2 S_{+}-\frac{1}{4}\sigma^2 S^{2}_{+}] \label{36}
	\end{equation}
then we obtain the thermodynamic quantities of interest as given by
	\begin{equation}
	\langle S_{+} \rangle=\frac{2}{\beta Z}\frac{\partial Z}{\partial \mu^{2}}, \langle H\rangle=\langle S^{2}_{+} \rangle=\frac{1}{ Z}\frac{\partial Z}{\partial \beta}+\mu^{2}\langle S_{+} \rangle
	\end{equation}
with the average value of the number of fermions and for bosons in the Universe being given by
	\begin{equation}
	\langle n_{F}\rangle=\frac{1}{ 2}(1-Z^{-1}),\langle n_{B}\rangle=\frac{\langle H \rangle}{E}-\langle n_{F}\rangle
	\end{equation}
As it was shown in the paper \cite{Kap} for small $\mu^{2}$ the relation between bosons and fermions in the universe may be presented as:$\frac{\langle n_{B}\rangle}{\langle n_{F}\rangle}=\coth \frac{\beta \sigma^{2}}{2}  $
and may take arbitrary predetermined values of this relation, depending on when the phase transition of the condensation of the bosonic part of the general representation of the wave function occurs. This observation indicates the reason for the bosonic asymmetry of the universe that is closely related to the baryonic asymmetry. This corresponds to the physical picture when the amount of overturning between different degrees of freedom is fixed in the system and the symmetry between the bosonic and fermionic subsystems is broken. This is the probable reason for the baryon asymmetry of the universe. At the same time, this leads to the usual spontaneous violation of symmetry that is necessary in the standard model. Now, if we remember that the supercharge by definition proposional to the density , it becomes obvious that the violation of the supersymmetry and the fixation of its relevant value leads to the birth of the matter. 
	
\textbf{Conclusion}
	
Clifford's algebraic formalism is proposed as a suitable method for describing the initial state of a vacuum with the possible birth of a fundamental field. This field should contain probable geometric characteristics and be fully equivalent to the traditional approach to quantum field theory with a richer structure. The approach makes it possible to explain the existence of supersymmetric properties of the original fundamental field, as well as the spontaneous breaking of symmetry between bosons and fermions in the universe. In addition, it makes it possible to explain the appearance of "dark matter" due to the influence of fields of tensor dimensions other than electromagnetic. In addition can "condense" part of the fields that do not have manifestations such as Bose or Ferm particles and the mixed state can manifest as "dark matter". Unfortunately, a rigorous mathematical proof of such an approach does not exist at the moment, but for purely physical reasons, such representations may favor better understanding of the scenario of the birth and evolution of the universe. After everything said above, it can be assumed that the energy of the initial vacuum state can be taken as zero. That is, due to the value of the coupling of the relevant field with unstable initial vacuum and the noise dispersion of such vacuum state determined all necessary initial parameter.

\textbf{Acknowledgment:} The author is sincerely grateful for the support of the Slovenian Research Agency (ARRS) through the programmer grant P1- 0099. This work was partially supported by Program of Fundamental Research of the Department of Physics and Astronomy of the National Academy of Sciences of Ukraine N 0120U101347	

\textbf{Data availability:}  Data sharing not applicable to this article as no data sets were generated or analyzed during the current study.No data associated in the manuscript.	

\textbf{Declarations}	Conflict of interest: The author declare no conflict of interest.

\end{document}